\documentclass[aps,floats]{revtex4}
\usepackage{amsmath,amssymb}
\usepackage{graphicx,epsfig}

\begin{document}
\bibliographystyle {plain}

\def\oppropto{\mathop{\propto}} 
\def\opsimeq{\mathop{\simeq}}
\def\opoverderline{\mathop{\overline}}
\def\operarrow{\mathop{\longrightarrow}}
\def\opsim{\mathop{\sim}} 
\def\opmin{\mathop{\min}} 
\def\opmax{\mathop{\max}} 

\def\fig#1#2{\includegraphics[height=#1]{#2}}
\def\figx#1#2{\includegraphics[width=#1]{#2}}



\title{ Localization transition in random L\'evy matrices : \\
multifractality of eigenvectors in the localized phase and at criticality }


\author{ C\'ecile Monthus }
 \affiliation{Institut de Physique Th\'{e}orique, 
Universit\'e Paris Saclay, CNRS, CEA,
91191 Gif-sur-Yvette, France}
 
\begin{abstract}

For random L\'evy matrices of size $N \times N$, where matrix elements are drawn with some heavy-tailed distribution $P(H_{ij}) \propto N^{-1} \vert H_{ij} \vert^{-1-\mu}$ with $0<\mu<2$ (infinite variance), there exists an extensive number of finite eigenvalues $E=O(1)$, while the maximal eigenvalue grows as $E_{max} \sim N^{\frac{1}{\mu}}$. Here we study the localization properties of the corresponding eigenvectors via some strong disorder perturbative expansion that remains consistent within the localized phase and that yields their Inverse Participation Ratios (I.P.R.)  $Y_q$ as a function of the continuous parameter $0<q<+\infty$. In the region $0<\mu<1$, we find that all eigenvectors are localized but display some multifractality : the IPR are finite above some threshold $q>q_c$ but diverge in the region $0<q<q_c$ near the origin. In the region $1<\mu<2$, only the sub-extensive fraction $N^{\frac{3}{2+\mu}}$ of the biggest eigenvalues corresponding to the region $\vert E \vert \geq N^{\frac{(\mu-1)}{\mu(2+\mu)}} $ remains localized, while the extensive number of other states of smaller energy are delocalized. For the extensive number of finite eigenvalues $E=O(1)$, the localization/delocalization transition thus takes place 
at the critical value $\mu_c=1$ corresponding to Cauchy matrices : the Inverse Participation Ratios $Y_q$ of the corresponding critical eigenstates follow the Strong-Multifractality Spectrum characterized by the generalized fractal dimensions $D^{criti}(q) = \frac{1-2q}{1-q} \theta (0 \leq q \leq \frac{1}{2})$, which has been found previously in various other Localization problems in spaces of effective infinite dimensionality.

\end{abstract}

\maketitle

\section{ Introduction}

In Anderson localization models \cite{anderson}, the localization properties of an eigenfunction $\psi (i)$
 defined on a finite sample containing $N$ sites can be measured by the well-known Inverse Participation Ratios (I.P.R.) 
(see the reviews \cite{janssenrevue,mirlinrevue} and references therein)
\begin{eqnarray}
Y_q(N)  \equiv 
\frac{ \sum_{i=1}^N  \vert \psi (i) \vert^{2q} }
{ \left[ \sum_{i=1}^N   \vert \psi (i) \vert^{2}  \right]^q }
\label{defipr}
\end{eqnarray}
where $q > 0 $ is a continuous index.
In the thermodynamic limit $N \to +\infty$, 
 they remain finite for exponentially localized eigenfunctions
\begin{eqnarray}
Y^{ exp \  loc}_{q>0}(N) \opsimeq_{N \to +\infty} O(1)
\label{iprexploc}
\end{eqnarray}
while for delocalized ergodic eigenfunctions where $\vert \psi (i) \vert^{2} \simeq \frac{1}{N} $
for any $i$, they display the power-laws
\begin{eqnarray}
Y^{ergodic}_q(N) \opsimeq_{N \to +\infty} N^{1-q}
\label{iprergodic}
\end{eqnarray}
i.e. they diverge $Y_q \to +\infty$ for $0<q<1$ and they converge towards zero $Y_q \to 0$
for $q>1$.
At the critical point between the two, 
the eigenfunctions display multifractality and
 the I.P.R. involve non-trivial exponents (see the reviews \cite{janssenrevue,mirlinrevue} and references therein)
\begin{eqnarray}
 Y_q^{multif}(N)   \opsimeq_{N \to \infty}  N^{ (1-q) D(q) }
\label{multif}
\end{eqnarray}u
with the generalized fractal dimensions $0 \leq D(q) \leq 1$,
where $D(q)=0$ corresponds to the localized behavior of Eq. \ref{iprexploc},
$D(q)=1$ corresponds to the ergodic behavior of Eq. \ref{iprergodic},
and intermediate values $0 < D(q) < 1$ correponds to multifractal eigenfunctions.

However there exists also models where localized eigenfunctions are not exponentially localized, but only power-law localized
with respect to the size $N$ of the Hilbert space. Important examples are
 Anderson models in finite dimension $d$ with power-law hoppings, 
  Anderson models on trees or on other spaces of effective infinite dimensionality where the Hilbert space grows exponentially with the linear length,
and some matrix models where the matrix elements are rescaled with some power of the size $N$ of the matrix.
Whenever the localization is power-law instead of exponential,
the IPR $Y_q$ cannot be expected to converge towards finite values for $q=0^+$,
but only above some threshold $q_c>0$ depending on the properties of the corresponding eigenstates
\begin{eqnarray}
Y^{ power \  loc}_{q>q_c}(N) \opsimeq_{N \to +\infty} O(1)
\label{iprpowerloc}
\end{eqnarray}
while the divergence in the region $0<q<q_c$ can be interpreted within the multifractal formalism of Eq. \ref{multif}
\begin{eqnarray}
Y^{ power \  loc}_{0<q<q_c}(N)    \opsimeq_{N \to \infty}  N^{ (1-q) D^{loc}(q) }
\label{locmultif}
\end{eqnarray}
as generalized dimensions decaying from $D^{loc}(q \to 0) \to 1 $ to $D^{loc}(q \to q_c) \to 0 $.
The goal of this paper is to analyse from this point of view the case of L\'evy matrices
where localized eigenvectors are expected to occur  
in some regions of the phase diagram parametrized by the L\'evy tail exponent $0<\mu<2$ and the energy 
\cite{cizeau,medina,soshnikov,birolitopeigen,burda,benarous,
auffinger_peche,metz,majumdar,auffinger_gui,bordenave,benaych,peche,biroli}.
In their pionneering paper, Bouchaud and Cizeau \cite{cizeau} have already stressed that 
the transition line for the IPR $Y_2$ for $q=2$ (denoted $Y$ in \cite{cizeau})
between the phases $Y_2=0(1)$ and $Y_2=0$ was different from the 
transition line for the IPR $Y_{1/2}$ for $q=1/2$ (denoted $\Upsilon$ in \cite{cizeau}, and which can be related to the properties
of the imaginary part of the self-energy \cite{cizeau})
between the phases $Y_{1/2}=0(1)$ and $Y_{1/2}=+\infty$ (see the phase diagram of Fig. 3 in \cite{cizeau}).
Here to obtain a fuller understanding of the localization properties of
eigenvectors of random L\'evy matrices, we wish to analyse the I.P.R. $Y_q$
for all values of the continuous index $q>0$ via some strong disorder approach based
on the strong inhomogeneities produced by L\'evy distributions.

The paper is organized as follows.
In Section \ref{sec_levy}, the L\'evy matrix model is described with its strong hierarchy of matrix elements.
In Section \ref{sec_square}, we analyse the statistical properties of the square of the L\'evy matrix.
In Section \ref{sec_per}, we study the strong disorder perturbative expansion for the square of the L\'evy matrix
and derive its domain of validity.
In Section \ref{sec_ipr}, we discuss the properties of the corresponding IPR as a function of the continuous index $q>0$.
Finally Section \ref{sec_cauchy} is devoted to the case of Cauchy matrices corresponding to the critical value $\mu_c=1$
for the localization/delocalization transitions of eigenvectors associated to finite eigenvalues.
Our conclusions are summarized in section \ref{sec_conclusion}.

\section{ Properties of random L\'evy matrices  } 

\label{sec_levy}

\subsection{ Model and notations }

The L\'evy matrix model introduced by Cizeau and Bouchaud \cite{cizeau}
is the set of $N \times N$ real symmetric matrices $H_{ij}=H_{ji}$ ,
where the entries $H_{i \leq j}$ are independent identical random variables
drawn with some heavy-tailed distribution of infinite variance.
For concreteness, it is convenient to consider the explicit simple example
of a pure power-law with some cut-off near the origin
\begin{eqnarray}
P(H_{ij}) dH_{ij} = \frac{ \mu H_N^{\mu}dH_{ij}}{ 2  \vert H_{ij} \vert^{1+\mu} } 
\theta \left( \vert H_{ij} \vert \geq H_N \right)
\label{levy}
\end{eqnarray}
where the important parameter is the tail exponent $0<\mu<2$, and where 
the cut-off is taken to depend on the size $N$ according to
\begin{eqnarray}
H_N = N^{-\frac{1}{\mu}}
\label{levy}
\end{eqnarray}

\subsection{ Strong hierarchy of the matrix elements  }

As stressed in \cite{cizeau}, this leads to a very strong hierarchy
in the matrix elements :

(i) the typical matrix element scales as
\begin{eqnarray}
H_{ij}^{typ} \propto H_N = N^{-\frac{1}{\mu}} 
\label{levytyp}
\end{eqnarray}

(ii) the largest matrix element $H_{i_0 j}$ seen by some site $i_0$ is however finite
\begin{eqnarray}
 \opmax_{1 \leq j \leq N} (H_{i_0 j} ) \propto H_N  N^{\frac{1}{\mu}}  \propto O(1)
\label{levymax}
\end{eqnarray}

(iii) the largest matrix element $H_{i j}$ in the whole matrix scales as
\begin{eqnarray}
 \opmax_{1 \leq i\leq j \leq N} (H_{i j} )\propto H_N  (N^2)^{\frac{1}{\mu}} \propto N^{\frac{1}{\mu}}
\label{levymaxtot}
\end{eqnarray}

\subsection{ Multifractal analysis of the matrix elements  }

To describe more precisely the inhomogeneities of the matrix elements,
it is useful to introduce the multifractal formalism
(see Appendix B of \cite{cizeau} even if the word 'multifractal' is not written explicitely) :
one considers the $N$ matrix elements $H_{i_0 j}$ seen by some site $i_0$,
and we may order them by their absolute values $\vert H_1 \vert > \vert H_2 \vert > .. $.
The scaling of the $n-th$ biggest absolute value $\vert H_n \vert $ scales as
\begin{eqnarray}
\int_{\vert H_n \vert}^{+\infty}  \frac{ \mu H_N^{\mu} dH}{ H^{1+\mu} } \simeq \frac{n}{N}
\label{levymaxninte}
\end{eqnarray}
yielding
\begin{eqnarray}
\vert H_n \vert \simeq H_N \left( \frac{N}{n} \right)^{\frac{1}{\mu}} = N^{-\frac{1}{\mu}}
 \left( \frac{N}{n} \right)^{\frac{1}{\mu}} = \left( \frac{1}{n} \right)^{\frac{1}{\mu}}
\label{levymaxnres}
\end{eqnarray}
With the change of variables $n=N^x$ with $0<x<1$, one obtains that 
the number of matrix elements scaling as $\vert H_{i_0 j} \vert \propto \vert N^{- \chi}$
scales as
\begin{eqnarray}
{\cal N} ( \vert H_{i_0 j} \vert \propto N^{-\chi} )
 \propto \int_0^1 dx N^x \delta (\chi-\frac{x}{\mu} )
\simeq  N^{ \mu \chi } \theta( 0 \leq \mu \chi \leq 1 )  \equiv N^{F(\chi)}
\label{levymultif}
\end{eqnarray}
corresponding to the linear multifractal spectrum of slope $\mu$
\begin{eqnarray}
F(\chi) =\mu \chi  \ \ \  \theta( 0 \leq  \chi \leq \frac{1}{\mu} )
\label{levyfchi}
\end{eqnarray}
The left boundary $\chi=0$ where the spectrum vanishes
\begin{eqnarray}
F(\chi=0) = 0
\label{levyfchi}
\end{eqnarray}
means that there are a finite number of matrix elements $H_{i_0 j}$ that are of order $O(1)$
as the maximum of Eq. \ref{levymax}.
The right boundary $ \chi_{typ} = \frac{1}{\mu} $ where the spectrum reaches its maximal value unity
\begin{eqnarray}
F(\chi_{typ}=\frac{1}{\mu}) = 1
\label{levyfchi}
\end{eqnarray}
means that there is an extensive number $O(N)$ of matrix elements
displaying the typical behavior $ N^{-\frac{1}{\mu}} $ of Eq. \ref{levytyp}.
The multifractal spectrum of Eq. \ref{levyfchi} is thus appropriate
to describe all the intermediate scales between the typical value and the maximal value.

\section{ Properties of the square of the L\'evy matrix  }

\label{sec_square}

One is interested into the spectral decomposition of the L\'evy matrix
\begin{eqnarray}
H = \sum_{i,j} H_{ij} \vert i > < j \vert  = \sum_{n=1}^N E_{n}  \vert \phi_n > < \phi_n \vert
\label{Hspectral}
\end{eqnarray}
into its $N$ eigenvalues $ E_{n} $ and the corresponding eigenvectors $\vert \phi_n >  $.

In this paper, we propose to focus on the square of the L\'evy matrix
\begin{eqnarray}
{\cal H} \equiv H^2  = \sum_{n=1}^N E^2_{n}  \vert \phi_n > < \phi_n \vert
\label{Hcarre}
\end{eqnarray}
which has the same eigenvectors $\vert \phi_n > $ and the related eigenvalues
\begin{eqnarray}
{\cal E}_n = E^2_{n} 
\label{Hcarreeigen}
\end{eqnarray}

The matrix elements of Eq. \ref{Hcarre} reads
\begin{eqnarray}
{\cal H}_{ij} \equiv <i \vert H^2 \vert j> = \sum_{k=1}^N H_{ik} H_{kj}
\label{Hcarreij}
\end{eqnarray}

\subsection{ Statistics of the diagonal matrix elements ${\cal H}_{ii} $  }

The diagonal element of Eq. \ref{Hcarreij}
\begin{eqnarray}
{\cal H}_{ii}  = \sum_{k=1}^N H_{ik}^2
\label{Hcarreij}
\end{eqnarray}
is a sum of $N$ independent positives variables 
\begin{eqnarray}
x_k \equiv H_{ik}^2
\label{xk}
\end{eqnarray}
By this change of variables, the distribution of Eq. \ref{levy} for $H_{ik}$
yields that the distribution $Q(x)$ of $x$ reads
\begin{eqnarray}
Q(x) = \frac{ \frac{\mu}{2} }{  N  x^{1+\frac{\mu}{2}} } 
\theta \left( x \geq N^{-\frac{\mu}{2}} \right)
\label{levyxk}
\end{eqnarray}
Since it displays a power-law tail involving the modified exponent $\frac{\mu}{2}<1$,
 the distribution of ${\cal H}_{ii} $
is given by the corresponding L\'evy stable distribution,
characterized by
 the Laplace transform
\begin{eqnarray}
\int_0^{+\infty} d {\cal H}_{ii} {\cal P}_{diag}({\cal H}_{ii}) e^{-t {\cal H}_{ii} } && =
\left[ \int_0^{+\infty} dx Q(x) e^{- t x} \right]^N = 
\left[ 1- \int_0^{+\infty} dx Q(x) (1-e^{- t x}) \right]^N
\nonumber \\
&& = e^{-\frac{\mu}{2} t^{\frac{\mu}{2}} \int_0^{+\infty} du \frac{ ( 1-e^{- u } ) }{   u^{1+\frac{\mu}{2}} } } 
\label{laplacediag}
\end{eqnarray}
So it displays the same power-law tail as Eq. \ref{levyxk}
 but without the prefactor $1/N$
\begin{eqnarray}
 {\cal P}_{diag}({\cal H}_{ii}) \opsimeq_{{\cal H}_{ii} \to +\infty}
 \frac{ \frac{\mu}{2} }{    {\cal H}_{ii}^{1+\frac{\mu}{2}} } 
\label{taildiag}
\end{eqnarray}
As a consequence, the typical value is finite
\begin{eqnarray}
 {\cal H}_{ii}^{typ} \propto O(1)
\label{diagtyp}
\end{eqnarray}
(which can be understood from the $O(1)$-scaling (Eq. \ref{levymax}) of the maximum term $H_{ik}^2 $ in Eq. \ref{Hcarreij}),
while the maximum value among the $N$ diagonal elements scales as
\begin{eqnarray}
\opmax_{1 \leq i \leq N} ( {\cal H}_{ii} ) \propto N^{\frac{2}{\mu}}
\label{diagmax}
\end{eqnarray}
(which can be understood from the scaling of the square of maximal matrix element in the whole matrix , see Eq. \ref{levymaxtot}).

For intermediate values, the multifractal analysis similar to Eq. \ref{levymultif}
yields that the number of diagonal elements ${\cal H}_{ii} $ scaling as  $ {\cal H}_{ii} \propto N^{\alpha}$ scales as
\begin{eqnarray}
{\cal N} ( {\cal H}_{ii} \propto N^{\alpha} )
 \propto   N^{ 1- \frac{\mu}{2} \alpha } \theta( 0 \leq  \alpha \leq \frac{2}{\mu} )  
\label{levymultifhii}
\end{eqnarray}
where the left boundary $\alpha=0$ describes the extensive number $O(N)$ of typical matrix elements
of order $O(1)$ of Eq. \ref{diagtyp}, while the right boundary $\alpha  = \frac{2}{\mu} $ 
describes the finite number of matrix elements governed by the maximal element of Eq. \ref{diagmax}.

Similarly, the joint distribution $  {\cal P}_{diag}({\cal H}_{11},{\cal H}_{22}, .., {\cal H}_{NN} )$ of the $N$ diagonal elements can be 
characterised by its Laplace transform
\begin{eqnarray}
&& \int_0^{+\infty} d {\cal H}_{11} .. \int_0^{+\infty} d {\cal H}_{NN}  {\cal P}_{diag}({\cal H}_{11}, .., {\cal H}_{NN} ) e^{- \sum_{i=1}^N t_i {\cal H}_{ii} }
\nonumber \\ && =
\left[ \prod_{i \leq j} \int_0^{+\infty} d H_{ij}  P(H_{ij} )\right] e^{- \sum_{i=1}^N t_i H_{ii} - \sum_{i<j} ( t_i +t_j) H_{ii} } 
\nonumber \\
&& \opsimeq_{N \to +\infty}  e^{-\frac{\mu}{2} \left[ \frac{1}{N} \sum_{i=1}^N  t_i^{\frac{\mu}{2}} +\frac{1}{N}   \sum_{i<j} ( t_i +t_j)^{\frac{\mu}{2}}  \right]  \int_0^{+\infty} du \frac{ ( 1-e^{- u } ) }{   u^{1+\frac{\mu}{2}}} } 
\label{laplacediagjointN}
\end{eqnarray}
In particular, the joint distribution of two diagonal elements reads
\begin{eqnarray}
 \int_0^{+\infty} d {\cal H}_{11}  \int_0^{+\infty} d {\cal H}_{22}  {\cal P}_{diag}({\cal H}_{11}, {\cal H}_{22} ) e^{-  t_1 {\cal H}_{11} - t_2 {\cal H}_{22}}
&& \opsimeq_{N \to +\infty}  e^{-\frac{\mu}{2} \left[ \left(1-\frac{1}{N} \right)  t_1^{\frac{\mu}{2}}+ \left(1-\frac{1}{N} \right)  t_2^{\frac{\mu}{2}}+\frac{1}{N}  ( t_1 +t_2)^{\frac{\mu}{2}}  \right]  \int_0^{+\infty} du \frac{ ( 1-e^{- u } ) }{   u^{1+\frac{\mu}{2}}} } 
\label{laplacediagjointN2}
\end{eqnarray}
so that the correlation term is only of order $1/N$ since it is due to the rare where the initial matrix element $H_{12}$ is of order $O(1)$.

\subsection{ Statistics of the off-diagonal matrix elements ${\cal H}_{i<j} $  }

The off-diagonal element of Eq. \ref{Hcarreij} for $i=1,j=2$ 
\begin{eqnarray}
{\cal H}_{12}  = \sum_{k=1}^N H_{1k} H_{2k} =  H_{12}( H_{11}+ H_{22} ) + \sum_{k=3}^N H_{1k} H_{2k}
\label{Hcarreijbis}
\end{eqnarray}
is the sum of $(N-2)$ independent variables for $k=3,..,N$
\begin{eqnarray}
y_k \equiv H_{1k} H_{2k}
\label{xk}
\end{eqnarray}
and of the special term $H_{12}( H_{11}+ H_{22} ) $ of the same form.
The distribution of $y_k$ reads from Eq. \ref{levy}
\begin{eqnarray}
R(y) = \int dH_1 P(H_1) \int dH_2 P(H_2)  \delta(y- H_1 H_2) 
=  \mu^2  
  \frac{ \ln \left(  \vert y \vert N^{\frac{2}{\mu}} \right) }{ 2 N^2  \vert y \vert^{1+\mu} }
\theta \left( \vert y \vert \geq N^{-\frac{2}{\mu}}\right)
\label{levyyk}
\end{eqnarray}
Since it is a symmetric heavy-tailed distribution of exponent $\mu$ modified by some logarithm,
one obtains that the off-diagonal element given by the sum of Eq. \ref{Hcarreijbis} is distributed
with a symmetric distribution with the Fourier transform
\begin{eqnarray}
\int_{-\infty}^{+\infty} d {\cal H}_{ij} {\cal P}_{off}({\cal H}_{ij}) e^{i \lambda t {\cal H}_{ij} } && 
\simeq 
\left[\int_{-\infty}^{+\infty} dy R(y) e^{i \lambda y}  \right]^{N} = 
e^{ - \frac{ \mu^2 }{N} \vert \lambda \vert^{\mu} \int_{0}^{+\infty} du 
  \frac{ \ln \left(   \frac{u}{\vert \lambda \vert}  \right) }{    u^{1+\mu} } 
 (1- \cos u ) }
\label{laplaceoff}
\end{eqnarray}
In particular, it displays the power-law tails of exponent $\mu$
\begin{eqnarray}
 {\cal P}_{off}({\cal H}_{ij}) \opsimeq_{{\cal H}_{ij} \to +\infty}
 \mu^2  
  \frac{ \ln \left(  \vert {\cal H}_{ij} \vert N^{\frac{1}{\mu}} \right) }{ 2 N  \vert {\cal H}_{ij} \vert^{1+\mu} }
\label{tailoff}
\end{eqnarray}
with an amplitude of order $1/N$.
As a consequence, the typical value scales as
\begin{eqnarray}
 {\cal H}_{ij}^{typ} \propto N^{-\frac{1}{\mu}}
\label{offtyp}
\end{eqnarray}
while the maximum value seen by some given site $i_0$ is finite
\begin{eqnarray}
\opmax_{j \ne i_0} ( {\cal H}_{i_0j} ) \propto O(1)
\label{offmax}
\end{eqnarray}

For intermediate values, the multifractal analysis 
thus yields exactly the same multifractal spectrum as Eq. \ref{levymultif} :
 the number of off-diagonal elements ${\cal H}_{i_0j} $ seen by $i_0$
scaling as  ${\cal H}_{i_0j}  \propto N^{-\chi}$ scales as
\begin{eqnarray}
{\cal N} ( \vert {\cal H}_{i_0 j} \vert \propto N^{-\chi} )
\simeq  N^{ \mu \chi } \theta( 0 \leq  \chi \leq \frac{1}{\mu} )  
\label{levymultifhi0j}
\end{eqnarray}

\section{ Strong Disorder perturbative expansion for the matrix ${\cal H}=H^2$  }

\label{sec_per}

\subsection{ First-order perturbation theory in the off-diagonal elements of $ {\cal H}=H^2$    }

Let us decompose the matrix $ {\cal H}=H^2$  into its diagonal and off-diagonal contributions
\begin{eqnarray}
{\cal H} && ={\cal H}_0+{\cal H}_1 \nonumber \\
{\cal H}_0 && = {\cal H}_{diag} = \sum_{i} {\cal H}_{ii} \vert i > < i \vert  \nonumber \\
{\cal H}_1 && ={\cal H}_{off} = \sum_{i<j} 
{\cal H}_{ij} \left( \vert i > < j \vert +  \vert j > < i \vert \right)
\label{Hperturbation}
\end{eqnarray}
Since the diagonal terms are typically of order $O(1)$,  (Eq \ref{diagtyp})
while the off-diagonal terms of typically of order $N^{-\frac{1}{\mu}} $ (Eq \ref{offtyp} ),
let us try to treat the off-diagonal part as a perturbation with respect to the diagonal part.

At order $0$, the eigenvectors of ${\cal H}_0 $ are completely localized on a single site
\begin{eqnarray}
  \vert \phi_i^{(0)} > = \vert i >
\label{phi0}
\end{eqnarray}
and the eigenvalues are given by the diagonal matrix elements
\begin{eqnarray}
  {\cal E}_i^{(0)} > = {\cal H}_{ii}
\label{eigen0}
\end{eqnarray}

At first order in perturbation theory, the eigenvalues remain unchanged
\begin{eqnarray}
  {\cal E}_i^{(0+1)} > = {\cal H}_{ii}
\label{eigen01}
\end{eqnarray}
while the eigenvectors become 
\begin{eqnarray}
\vert \phi_{i_0}^{(0+1)} > && 
  = \vert i_0 > + \sum_{j \ne i_0} \vert j >  R_{i_0j}
\label{phi1}
\end{eqnarray}
in terms of the hybridization ratios
\begin{eqnarray}
R_{i_0j}  && \equiv 
   \frac{{\cal H} _{i_0j}  }{ {\cal H}_{i_0i_0}-{\cal H}_{jj}  }
\label{ri0j}
\end{eqnarray}

The corresponding Inverse Participation Ratios of Eq. \ref{defipr} read
\begin{eqnarray}
Y_q(i_0)= \frac{ \sum_{j}
 \vert \phi _{i_0}^{(0+1)}(j) \vert^{2q} }
{ \left[  \sum_{j} \vert \phi_{i_0}^{(0+1)} (j) \vert^{2}  \right]^q }
= \frac{1+\Sigma_q(i_0)}{ (1+\Sigma_1(i_0) )^q }
  \label{yq}
\end{eqnarray} 
in terms of the sums over the $(N-1)$ hybridization ratios
\begin{eqnarray}
\Sigma_q(i_0)  \equiv  \sum_{j \ne i_0} \vert R_{i_0j} \vert^{2q}
  \label{sigmaq}
\end{eqnarray}

The localization properties of the eigenvector $\vert \phi_{i_0}^{(0+1)} > $
depend on its energy ${\cal H}_{i_0i_0} $  :
in the following sections, we consider the multifractal parametrization of Eq. \ref{levymultifhii}
\begin{eqnarray}
 {\cal H}_{i_0i_0} \propto N^{\alpha_0} 
\label{alpha0}
\end{eqnarray}
where $ 0 \leq  \alpha_0 \leq \frac{2}{\mu} $.
Note that there are $ N^{ 1- \frac{\mu}{2} \alpha_0 }$ of such states (Eq. \ref{levymultifhii}).

To determine whether the perturbative expansion in the off-diagonal elements makes sense,
we need to analyse the statistical properties of the hybridization ratios of Eq. \ref{ri0j}

\subsection{ Typical hybridization ratio }

The typical hybridization ratio to a state of energy Eq. \ref{alpha0}
can be estimated from the two typical values of Eqs \ref{diagtyp} and \ref{offtyp}
for ${\cal H}_{jj}^{typ} $ and ${\cal H} _{i_0j}^{typ} $ respectively
\begin{eqnarray}
R_{i_0j}^{typ}  && \simeq 
   \frac{{\cal H} _{i_0j}^{typ}  }{N^{\alpha_0}  -{\cal H}_{jj}^{typ}  }  
 \propto 
\frac{ N^{-\frac{1}{\mu}}}{N^{\alpha_0} -O(1) } \propto  N^{-\frac{1}{\mu}-\alpha_0 }
\label{ri0jtyp}
\end{eqnarray}
It decays with $N$ for any value $ 0 \leq  \alpha_0 \leq \frac{2}{\mu} $,
so for a typical term, the perturbative expansion makes sense.
However, we now need to consider non-typical index $j$ that can produce
anomalously larger hybridization ratio, either via large off-diagonal coupling,
or via small difference between diagonal energies.

\subsection{ Hybridization Ratio with the state having the biggest off-diagonal coupling }

The site $j_{max}$ having the biggest off-diagonal coupling with $i_0$ remains finite (Eq \ref{offmax})
\begin{eqnarray}
{\cal H}_{i_0j_{max}} =  \opmax_{j \ne i_0} ( {\cal H}_{i_0j} ) \propto O(1)
\label{offjmax}
\end{eqnarray}
and yields the hybridization ratio
\begin{eqnarray}
R_{i_0j_{max}}  && \simeq
   \frac{{\cal H} _{i_0j_{max}}  }{N^{\alpha_0}  -{\cal H}_{jj}^{typ}  }  
 \propto 
\frac{ O(1) }{N^{\alpha_0} -O(1) } \propto  N^{-\alpha_0 }
\label{ri0offmax}
\end{eqnarray}
It decays with $N$ for $\alpha_0>0$, while it remains finite $R_{i_0j_{max}}=O(1) $
for $\alpha_0=0$ corresponding to finite energy ${\cal H}_{i_0i_0} \propto O(1)$ 
(Eq \ref{alpha0}).

\subsection{ Hybridization Ratio with the consecutive energy level }

Since there are $ N^{ 1- \frac{\mu}{2} \alpha_0 }$ other states with the same scaling 
$H_{jj} \propto N^{\alpha_0} $ as $  H_{i_0i_0}$ in Eq. \ref{alpha0},
the level spacing in this region of the spectrum scales as
\begin{eqnarray}
 \vert {\cal H}_{i_0i_0}- {\cal H}_{next} \vert \propto \frac{ N^{\alpha_0} }{ N^{ 1- \frac{\mu}{2} \alpha_0 }}
= N^{ -1 + \left(1+ \frac{\mu}{2}\right) \alpha_0 }
\label{levelspacingalpha0}
\end{eqnarray}
In particular, one recovers the standard $N^{-1}$ level spacing in the region of finite energy 
corresponding to $\alpha_0=0$, and the level spacing $N^{\frac{2}{\mu}}$ in the region near the maximal eigenvalue
corresponding to $\alpha_0=\frac{2}{\mu}$.

The hybridization ratio with the consecutive energy level then scales as
\begin{eqnarray}
R_{i_0j}^{next}  && \equiv 
   \frac{{\cal H} _{i_0j}^{typ}  }{{\cal H}_{i_0i_0}  -{\cal H}_{next}  }  
 \propto 
\frac{ N^{-\frac{1}{\mu}}}{  N^{ -1 + \left(1+ \frac{\mu}{2}\right) \alpha_0 }  } \propto 
 N^{ 1-\frac{1}{\mu} - \left(1+ \frac{\mu}{2}\right) \alpha_0 }
\label{rationext}
\end{eqnarray}

So here one needs to distinguich two regions :

(i) For $0<\mu<1$, the hybridization ratio of Eq. \ref{rationext} decays with $N$
for arbitrary energy $ 0 \leq  \alpha_0 \leq \frac{2}{\mu} $ in Eq. \ref{alpha0}.
So from this criterion based on the vanishing hybridization ratio with the consecutive energy level,
one obtains that the perturbative expansion is consistent for all energies.

(ii) For $1<\mu<2$,  one needs to introduce the threshold
\begin{eqnarray}
 \alpha_c(\mu) \equiv \frac{\mu-1}{\mu \left(1+ \frac{\mu}{2}\right)} 
\label{alphac}
\end{eqnarray}
going from $ \alpha_c(\mu=1)=0 $ to $ \alpha_c(\mu=2)=1/4 $.
The hybridization ratio of Eq. \ref{rationext} 
\begin{eqnarray}
R_{i_0j}^{next}  && \propto 
 N^{  \left(1+ \frac{\mu}{2}\right) ( \alpha_c(\mu)  - \alpha_0) }
\label{rationext}
\end{eqnarray}
decays with $N$
only for sufficiently large energy defined by the energy exponent region
 \begin{eqnarray}
\alpha_c(\mu)<\alpha_0 \leq \frac{2}{\mu}
\label{validity}
\end{eqnarray}
So here the perturbative expansion is consistent only for these high-energy levels,
whose number scales as
\begin{eqnarray}
\int_{\alpha_c(\mu)}^{\frac{2}{\mu}} d \alpha_0 N^{1- \frac{2}{\mu}\alpha_0} \simeq 
N^{1- \frac{\mu}{2}\alpha_c(\mu)} = N^{ \frac{3}{ 2+\mu}  }
\label{numberloc}
\end{eqnarray}
For $\mu \to 1$, one recovers the scaling of $N$ states to match the region (i),
while for $\mu \to 2$, the perturbative expansion makes sence only for the $N^{\frac{3}{4}}$ highest energy states.

For the other states of sufficiently small energy defined by the energy exponent region $0 \leq \alpha_0< \alpha_c(\mu) $,
 the hybridization ratio
with the consecutive level of Eq. \ref{rationext} diverges and will thus produces a complete mixing
with a diverging number of nearby-energy states : our conclusion is thus that the eigenstates become delocalized.
A very interesting issue is whether these delocalized states are ergodic (Eq. \ref{iprergodic}) or non-ergodic
with some non-trivial multifractal spectrum (Eq. \ref{multif}). The existence of delocalized non-ergodic states
has been discussed in various models \cite{levitov,gornyi_fock,vadim,us_mblaoki,us_cayley,biroli_nonergo,luca,gornyi,kravtsov_nonergomf,c_mblrgeigen}
but remains very controversial even for Random-Regular-Graphs as shown by the two very recent studies with opposite conclusions
\cite{mirlin_ergo,altshuler_nonergo}. Since the strong disorder perturbation approach described above is not consistent anymore
in the delocalized region $0 \leq \alpha_0< \alpha_c(\mu) $, we cannot address this issue here and leave it for future work
since it requires a completely different approach.

\section{ Analysis of the Inverse Participation Ratios $Y_q$  }

\label{sec_ipr}

In this section, we analyse the IPR $Y_q$ of Eq. \ref{yq}
for the states where the perturbative expansion described in the previous section makes sense.
So we need to study the statistical properties of the sums of Eq. \ref{sigmaq}
\begin{eqnarray}
\Sigma_q(i_0)  \equiv  \sum_{j \ne i_0} w_q(j) 
  \label{sigmaqwq}
\end{eqnarray} 
with the notations
\begin{eqnarray}
w_q(j) && \equiv u_q(j) v_q(j)
\nonumber \\
u_q(j)  &&\equiv \vert  {\cal H}_{i_0j} \vert ^{2q}
\nonumber \\
v_q(j)  &&\equiv \vert {\cal H}_{jj}-{\cal H}_{i_0i_0} \vert ^{-2q}
  \label{wuv}
\end{eqnarray} 
So let us analyse the probability distributions of $u_q$, of $v_q$ and of $w_q$.

\subsection{ Probability distribution of $u_q(j) \equiv \vert  {\cal H}_{i_0j} \vert ^{2q}$  }

The change of variable between the off-diagonal element ${\cal H}_{i_0j} $ 
distributed with ${\cal P}_{off} ({\cal H}_{i_0j}) $ of Eqs \ref{laplaceoff} and \ref{tailoff}
and the new positive variable 
\begin{eqnarray}
u_q(j) \equiv \vert  {\cal H}_{i_0j} \vert ^{2q}
\label{uq}
\end{eqnarray}
yields the probability distribution $U_q$ displaying the tail of exponent $\frac{\mu}{2q}  $
\begin{eqnarray}
U_q(u_q) \opsimeq_{u_q \to +\infty}  \left( \frac{\mu}{2q} \right)^2  
  \frac{ \ln u_q  }{   N  u_q^{1+\frac{\mu}{2q} } } 
\label{distriuq}
\end{eqnarray}
In particular, the typical value scales as
\begin{eqnarray}
 u_q^{typ} \propto N^{-\frac{2q}{\mu}}
\label{uqtyp}
\end{eqnarray}
while the maximum value seen by some given site $i_0$ is finite
\begin{eqnarray}
\opmax_{j \ne i_0} ( u_q(j) ) \propto O(1)
\label{uqmax}
\end{eqnarray}
For intermediate values, the multifractal analysis
yields that the number of $u_q(j) $ scaling as  $u_q(j)  \propto N^{-\chi}$ scales as
\begin{eqnarray}
{\cal N} ( u_q(j)  \propto N^{-\chi} )
 \propto   N^{ \frac{\mu}{2q} \chi } \theta( 0 \leq \frac{\mu}{2q} \chi \leq 1 )  
\label{levymultifU}
\end{eqnarray}

\subsection{Probability distribution of
 $v_q(j) \equiv \vert {\cal H}_{jj}-{\cal H}_{i_0i_0} \vert ^{-2q}$  }

Here ${\cal H}_{i_0i_0}=N^{\alpha_0} $ is considered as fixed and given (Eq \ref{alpha0}).
The variable
\begin{eqnarray}
v_q(j) \equiv \vert  {\cal H}_{i_0i_0}- H_{jj}   \vert^{-2 q}
\label{vq}
\end{eqnarray}
has for typical value
\begin{eqnarray}
 v_q^{typ} \propto \vert  {\cal H}_{i_0i_0}- H_{jj}^{typ}   \vert^{-2 q} = \vert N^{\alpha_0}-O(1)  \vert^{-2 q} = N^{-2q \alpha_0}
\label{vqtyp}
\end{eqnarray}

The change of variable between the diagonal element ${\cal H}_{jj} $ 
distributed with ${\cal P}_{diag} ({\cal H}_{jj}) $ of Eqs \ref{laplacediag} and \ref{taildiag} 
and $v_q(j)$
yields that the probability distribution $V_q$ displays the power-law tail 
\begin{eqnarray}
V_q^{sing}(v_q) \opsimeq_{v_q \to +\infty} \frac{{\cal P}_{diag}({\cal H}_{i_0i_0})  }{  2q v_q^{1+\frac{1}{2q}}}
&& \opsimeq_{v_q \to +\infty} \frac{ \frac{\mu}{2}  }{N^{\alpha_0 (1+\frac{\mu}{2})}  2q v_q^{1+\frac{1}{2q}}}
\label{distrivq}
\end{eqnarray}
In particular, the corresponding maximum value seen by some given site $i_0$ scales as
\begin{eqnarray}
v_q^{max} = \opmax_{j \ne i_0} ( v_q(j) ) \propto N^{2q \left( 1-\alpha_0 (1+\frac{\mu}{2}) \right) }
\label{vqmax}
\end{eqnarray}
in agreement with the behavior of the level spacing of Eq. \ref{levelspacingalpha0} 
in this region of the spectrum.

Here for $\alpha_0>0$, the typical value of Eq. \ref{vqtyp} is not contained in the rare-event tail of Eq. \ref{distrivq}
that would correspond to the smaller characteristic scale $ N^{-2q \alpha_0(1+\frac{\mu}{2})} $,
so we write the probability distribution $V_q$ as a sum of two contributions :
 one regular contribution involving the typical value and the singular contribution of Eq. \ref{distrivq}
\begin{eqnarray}
V_q(v_q) \simeq \frac{1}{ v_q^{typ}} V_q^{reg}\left(\frac{v_q}{ v_q^{typ}}\right) +V_q^{sing}(v_q) 
\label{distrivqfull}
\end{eqnarray}

\subsection{Probability distribution of
 $w_q(j) \equiv \vert \frac{ {\cal H}_{i_0j}  }{{\cal H}_{i_0i_0}-{\cal H}_{jj}  } \vert^{2q}$  }

Let us now consider the product
\begin{eqnarray}
w_q(j) \equiv u_q(j) v_q(j)
 =\vert  {\cal H}_{i_0j} \vert ^{2q} \vert H_{jj} - {\cal H}_{i_0i_0} \vert^{-2 q}
\label{wq}
\end{eqnarray}
Its typical value scales as (Eqs \ref{uqtyp} and \ref{vqtyp})
\begin{eqnarray}
w_q^{typ} \equiv u_q^{typ} v_q^{typ} \propto  N^{-\frac{2q}{\mu}}  N^{-2q \alpha_0}
 = N^{-2q \left(\frac{1}{\mu}+\alpha_0 \right)}
\label{wqtyp}
\end{eqnarray}
To evaluate its maximal value when drawing $N$ variables,
let us compare the first possibility (Eq. \ref{uqmax})
\begin{eqnarray}
w_q^{max(1)} && \equiv u_q^{max} v_q^{typ} \propto    N^{-2q \alpha_0}
\label{wqmaxposs1}
\end{eqnarray}
in agreement with the biggest off-diagonal coupling of Eq. \ref{ri0offmax},
with the second possibility (Eq. \ref{vqmax})
\begin{eqnarray}
w_q^{max(2) } && \equiv u_q^{typ} v_q^{max} \propto
N^{2q \left( 1- \frac{1}{\mu} -\alpha_0 (1+\frac{\mu}{2}) \right) } 
\label{wqmaxposs2}
\end{eqnarray}
that corresponds to the smallest energy difference of Eq. \ref{rationext}.

Let us now turn to the probability distribution of $w_q$ :
the probability distribution of $u_q$ displays a power-law tail of exponent $(1+\frac{\mu}{2q})$ (Eq. \ref{distriuq}), 
while the distribution of $v_q$ displays a power-law tail of exponent $(1+\frac{1}{2q})$ (Eq. \ref{distriuq}), 
so the probability distribution of $w_q$ will inherit
a power-law tail with the smallest exponent between the two, 
so the result depend on the position of $\mu$
with respect to the value $\mu=1$.

\subsection{ Case $0<\mu<1$  }

For $0<\mu<1$, the tail exponent $\frac{\mu}{2q}  $ of Eq. \ref{distriuq}
is smaller than the tail exponent $\frac{1}{2q}  $ of Eq. \ref{distrivq},
so that the distribution of $w_q$ will inherit the tail of Eq. \ref{distriuq}
with an amplitude modified by the finite non-integer moment
 $\overline{ v_q^{\frac{\mu}{2q}}}$ that can be evaluated from Eq. \ref{distrivqfull}
\begin{eqnarray}
\overline{ v_q^{\frac{\mu}{2q}}} = \int dv_q  v_q^{\frac{\mu}{2q}}  V_q(v_q) \simeq
(v_q^{typ})^{\frac{\mu}{2q}} +     \frac{ (cst)  }{N^{\alpha_0 (1+\frac{\mu}{2})}}
= {\rm max} (N^{- \mu \alpha_0}  , N^{-\alpha_0 (1+\frac{\mu}{2})}) = N^{- \mu \alpha_0}
\label{momentvqfull}
\end{eqnarray}
so that one obtains the tail
\begin{eqnarray}
W_q(w_q) \opsimeq_{w_q \to +\infty}  \left( \frac{\mu}{2q} \right)^2  (v_q^{typ})^{\frac{\mu}{2q}}
  \frac{ \ln w_q  }{   N  w_q^{1+\frac{\mu}{2q} } } 
=  \left( \frac{\mu}{2q} \right)^2 
  \frac{ \ln w_q  }{   N^{1+\mu \alpha_0}  w_q^{1+\frac{\mu}{2q} } } 
\label{distriwqsmallmu}
\end{eqnarray}
The maximum value among $N$ scales as
\begin{eqnarray}
w_q^{max} && \simeq   N^{-2q \alpha_0}
\label{wqmaxmusmall}
\end{eqnarray}
in agreement Eq. \ref{wqmaxposs1}, while the typical value agrees with Eq. \ref{wqtyp}.

Now we are interested into the sum $\Sigma_q$ of $(N-1)$ of such positives variables (Eq. \ref{sigmaqwq})

(i) for sufficiently big indices $q>\frac{\mu}{2} $, the exponent $\frac{\mu}{2q}  $ in Eq. \ref{distriwqsmallmu}
is smaller than unity : the probability distribution $\Pi_q$ of $\Sigma_q$ is then a L\'evy stable law of exponent $\frac{\mu}{2q}  $
with the tail obtained from Eq. \ref{distriwqsmallmu}
\begin{eqnarray}
\Pi_q(\Sigma_q) \opsimeq_{\Sigma_q \to +\infty}  \left( \frac{\mu}{2q} \right)^2  
  \frac{ \ln \Sigma_q  }{   N^{\mu \alpha_0}   \Sigma_q^{1+\frac{\mu}{2q} } } 
\label{distrisigmaqsmallmu}
\end{eqnarray}
so that its typical value displays the same scaling as the maximal term of Eq. \ref{wqmaxmusmall}
\begin{eqnarray}
\Sigma_q^{typ} && \simeq   N^{-2q \alpha_0}
\label{sigmaqbigmusmall}
\end{eqnarray}
It remains finite for finite energy corresponding to $\alpha_0=0$,
while it converges towards zero for the other energy exponents $0<\alpha_0<\frac{2}{\mu}$.

(ii) for smaller indices $0<q<\frac{\mu}{2} $, the exponent $\frac{\mu}{2q}  $ in Eq. \ref{distriwqsmallmu}
is bigger than unity :
the average value of $w_q$ does not diverge
and is scales as the typical value $w_q^{typ}$ of Eq. \ref{wqtyp}
\begin{eqnarray}
\overline{ w_q} \simeq w_q^{typ}
\propto    N^{-2q \left(\frac{1}{\mu}+\alpha_0 \right)}
\label{wqav}
\end{eqnarray}
so that the sum scales as
\begin{eqnarray}
\Sigma_q \simeq N \overline{ w_q} \propto   N^{1-2q \left(\frac{1}{\mu}+\alpha_0 \right)} 
\label{sqtyp}
\end{eqnarray}
In the  thermodynamic limit $N \to +\infty $, it diverges for
$q<q_c(\mu,\alpha_0)$, remains finite for $q=q_c(\mu,\alpha_0)$,
 and converges towards zero for  $q_c(\mu,\alpha_0)<q<\frac{\mu}{2}$,
in terms of the threshold
\begin{eqnarray}
q_c(\mu,\alpha_0) \equiv \frac{1}{2 \left(\frac{1}{\mu}+\alpha_0 \right)}
\label{qc}
\end{eqnarray}

The particular value $q=1$ belongs to the case (i) and thus the sum $\Sigma_1$
appearing in the denominator of Eq. \ref{ipr} does not diverge.
As a consequence, the behavior of the IPR $Y_q$ is directly 
determined by the convergence or divergence of $\Sigma_q$ :
from the above results, one thus obtains that
 the generalized dimensions of Eq. \ref{multif} read
\begin{eqnarray}
D^{(0<\mu<1;0 \leq \alpha_0 \leq \frac{2}{\mu})}(q) && =  0   \  \ \  \ \ \  \ \ \ \ \ \  \ \ \ \ \  \ \ \ {\rm for } \ \ \ q> q_c(\mu,\alpha_0)
\nonumber \\
D^{(0<\mu<1;0 \leq \alpha_0 \leq \frac{2}{\mu})}(q) && =  \frac{1- \frac{q }{q_c(\mu,\alpha_0)}}{1-q}  \ \ \  \ \ \ {\rm for } \ \ \ 0<q< q_c(\mu,\alpha_0)
\label{multifmusmall}
\end{eqnarray}
which corresponds to the following linear multifractal spectrum :
the number of hybridization ratios scaling as $R_{i_0j} \propto N^{-\frac{\gamma}{2}} $ scales as 
\begin{eqnarray}
{\cal N}^{(0<\mu<1;0 \leq \alpha_0 \leq \frac{2}{\mu})} ( R_{i_0j} \propto N^{-\frac{\gamma}{2}} )
&& \propto   N^{\frac{\gamma}{\gamma_0(\mu,\alpha_0)}}  \theta (0 \leq \gamma \leq \gamma_0(\mu,\alpha_0))
\label{levymultifR}
\end{eqnarray}
with the typical exponent
\begin{eqnarray}
\gamma_0(\mu,\alpha_0) \equiv \frac{1}{q_c(\mu,\alpha_0)} =2 \left(\frac{1}{\mu}+\alpha_0 \right)
\label{levymultifRmusmall}
\end{eqnarray}
governing an extensive number of ratios,
while a finite number $O(1)$ of ratios remain finite with the value $\gamma=0$.
Note that this type of linear spectrum for the Localized phase
has already been found for an Anderson Localization matrix model in \cite{kravtsov_nonergomf}
and for the Many-Body-Localization models in \cite{c_mblperturb,c_mblrgeigen}.

\subsection{ Case $1<\mu<2$  }

For $1<\mu<2$, the tail exponent $\frac{1}{2q}  $ of Eq. \ref{distrivq}
is smaller than the tail exponent $\frac{\mu}{2q}  $ of Eq. \ref{distriuq},
so that the distribution of $w_q$ will inherit the tail of Eq. \ref{distrivq}
with an amplitude modified by the non-integer moment
 $\overline{ u_q^{\frac{1}{2q}}}$ behaving as $(u_q^{typ})^{\frac{1}{2q}}=N^{-\frac{1}{\mu}} $
\begin{eqnarray}
W_q^{sing}(w_q) \opsimeq_{w_q \to +\infty}
 \frac{ \frac{\mu}{2} }{N^{\frac{1}{\mu}+\alpha_0 (1+\frac{\mu}{2})}  2q w_q^{1+\frac{1}{2q}}}
\label{distriwqsbigmu}
\end{eqnarray}
so that the corresponding maximum value among $N$ scales as
\begin{eqnarray}
w_q^{max} && \simeq  
N^{2q \left( 1- \frac{1}{\mu} -\alpha_0 (1+\frac{\mu}{2}) \right) }
\label{wqmaxmubig}
\end{eqnarray}
in agreement with Eq. \ref{wqmaxposs2}.
Together with the regular contribution, the full distribution reads
\begin{eqnarray}
W_q(w_q) \simeq \frac{1}{ w_q^{typ}} W_q^{reg}\left(\frac{w_q}{ w_q^{typ}}\right) +W_q^{sing}(w_q) 
\label{distriwqfull}
\end{eqnarray}

Now we are interested into the sum $\Sigma_q$ of $(N-1)$ of such positives variables (Eq \ref{sigmaqwq}).
The regular part of Eq. \ref{distriwqfull} yields the scaling
\begin{eqnarray}
\Sigma_q^{reg} \propto  N w_q^{typ} =  N^{1-2q \left(\frac{1}{\mu}+\alpha_0 \right)}
\label{sigmaqreg}
\end{eqnarray}

Let us now discuss the scaling produced by the singular part of Eq. \ref{distriwqfull}

(i) for big indices $q>\frac{1}{2} $, 
the exponent $\frac{1}{2q}  $ in Eq. \ref{distriwqsbigmu}
is smaller than unity : so the singular part of the probability distribution $\Pi_q$ of $\Sigma_q$ 
displays the tail inherited from Eq.  \ref{distriwqsbigmu}
\begin{eqnarray}
\Pi_q^{sing}(\Sigma_q) \opsimeq_{\Sigma_q \to +\infty}  
 \frac{ \frac{\mu}{2} }{N^{\frac{1}{\mu}-1+\alpha_0 (1+\frac{\mu}{2})}  2q\Sigma_q^{1+\frac{1}{2q}}}
\label{distrisigmaqbigmu}
\end{eqnarray}
leading to the scaling (with the notation of Eq. \ref{alphac})
\begin{eqnarray}
\Sigma_q^{sing} \propto  N^{2q \left( 1- \frac{1}{\mu} -\alpha_0 (1+\frac{\mu}{2}) \right) }
\label{sigmaqsingbigmuqbig}
\end{eqnarray}
in agreement with the maximal term of Eq \ref{wqmaxmubig}.
This converges in the region $\alpha_c(\mu) \leq \alpha_0 \leq \frac{2}{\mu}$
where the perturbative expansion makes sense (Eq. \ref{validity}).

Putting the contributions of Eqs \ref{sigmaqreg} and \ref{sigmaqsingbigmuqbig} together,
the final result for the scaling in the region $q>\frac{1}{2}  $ reads
\begin{eqnarray}
\Sigma_{q>\frac{1}{2}} \propto  {\rm max} (  N^{1-2q \left(\frac{1}{\mu}+\alpha_0 \right)} ,  N^{- 2q (1+\frac{\mu}{2})( \alpha_0 - \alpha_c(\mu) ) } )
\label{sigmaqtypqbig}
\end{eqnarray}

(ii) for small indices $0<q<\frac{1}{2} $, 
the exponent $\frac{1}{2q}  $ in Eq. \ref{distriwqsbigmu}
is bigger than unity :
the average value of $w_q$ can be then obtained from Eq. \ref{distriwqfull}
\begin{eqnarray}
\overline{ w_q} && \simeq  \int dw_q w_q W_q(w_q) \propto w_q^{typ}+ 
+ \frac{ (cst) }{N^{\frac{1}{\mu}+\alpha_0 (1+\frac{\mu}{2}) }}
\nonumber \\
 && \propto {\rm max} (   N^{-2q \left(\frac{1}{\mu}+\alpha_0 \right)} ;N^{-\frac{1}{\mu}-\alpha_0 (1+\frac{\mu}{2})}  ) 
\label{wqavbis}
\end{eqnarray}
so that the sum scales as
\begin{eqnarray}
\Sigma_{q<\frac{1}{2}}  \simeq N \overline{ w_q} && \propto   {\rm max} (   N^{1-2q \left(\frac{1}{\mu}+\alpha_0 \right)} ;N^{1-\frac{1}{\mu}-\alpha_0 (1+\frac{\mu}{2})}  ) 
\nonumber \\ 
&& \propto   {\rm max} (   N^{1-2q \left(\frac{1}{\mu}+\alpha_0 \right)} ;N^{- (1+\frac{\mu}{2} ) (\alpha_0-\alpha_c(\mu) )} )
\label{sqtypqsmall}
\end{eqnarray}

In the region of validity of the perturbative expansion (Eq. \ref{validity}), the second exponent in Eqs \ref{sigmaqtypqbig}
and \ref{sqtypqsmall}  is always negative,
so the only possibility of divergence of $\Sigma_q$ occurs when the exponent of the first factor is positive,
i.e. for sufficiently small indices  $q<q_c(\mu,\alpha_0) $ in terms of the threshold $q_c(\mu,\alpha_0) $ introduced in Eq. \ref{qc}.
Since $q_c(\mu,\alpha_0)<1 $, the sum $\Sigma_1$ for $q=1$ is finite, and the behavior of the IPR $Y_q$ is directly 
determined by the convergence or divergence of $\Sigma_q$.
As a consequence, the generalized dimensions of Eq. \ref{multif} reads
\begin{eqnarray}
D^{(1<\mu<2;\alpha_c(\mu) \leq \alpha_0 \leq \frac{2}{\mu})}(q) && =  0   \  \ \  \ \ \  \ \ \ \ \ \  \ \ \ \ \  \ \ \ {\rm for } \ \ \ q> q_c(\mu,\alpha_0)
\nonumber \\
D^{(1<\mu<2;\alpha_c(\mu) \leq \alpha_0 \leq \frac{2}{\mu})}(q) && =  \frac{1- \frac{q }{q_c(\mu,\alpha_0)}}{1-q}  \ \ \  \ \ \ {\rm for } \ \ \ 0<q< q_c(\mu,\alpha_0)
\label{multifmubig}
\end{eqnarray}
which corresponds to the linear multifractal spectrum
\begin{eqnarray}
{\cal N}^{(1<\mu<2;\alpha_c(\mu) \leq \alpha_0 \leq \frac{2}{\mu})} ( R_{i_0j} \propto N^{-\frac{\gamma}{2}} )
&& \propto   N^{\frac{\gamma}{\gamma_0}}  \theta (0 \leq \gamma \leq \gamma_0)
\label{levymultifRbig}
\end{eqnarray}
with the typical exponent
\begin{eqnarray}
\gamma_0(\mu,\alpha_0)  \equiv \frac{1}{q_c(\mu,\alpha_0)} =2 \left(\frac{1}{\mu}+\alpha_0 \right)
\label{levymultifRmubig}
\end{eqnarray}

\section{ The critical case of Cauchy matrices $\mu=1$ for finite energies $\alpha_0=0$}

\label{sec_cauchy}

In the region of finite energies corresponding to the exponent $\alpha_0$ (Eq \ref{alpha0}) and containing
the extensive $O(N)$ number of eigenvalues, the strong disorder approach of section \ref{sec_per}
points towards the critical value $\mu_c=1$ between localized states for $0<\mu<1$
and delocalized states for $1<\mu<2$. It seems thus useful to reconsider the various steps of the calculations for the special case of Cauchy matrices,
where the matrix elements are drawn with the Cauchy distribution (instead of Eq. \ref{levy})
\begin{eqnarray}
C(H_{ij})  = \frac{ a_N }{ \pi ( a_N^2+  H_{ij} ^2 )} 
\label{cauchyy}
\end{eqnarray}
where the scale $a_N$ is chosen to match the asymptotic behavior of Eq. \ref{levy}
\begin{eqnarray}
a_N = \frac{\pi}{2N}
\label{ancauchy}
\end{eqnarray}
in order to facilitate the comparison between formulas.

\subsection{ Properties of the square of the Cauchy matrix ${\cal H}=H^2$  }

The probability distribution of the positive variable $x_k \equiv H_{ik}^2$ (Eq. \ref{xk}) reads (instead of Eq. \ref{levyxk})
\begin{eqnarray}
Q(x) =  \frac{ a_N }{ \sqrt{x}  \pi ( a_N^2+  x )} \opsimeq_{x \to +\infty}  \frac{ 1 }{    2N x^{\frac{3}{2}} }
\label{cauchyxk}
\end{eqnarray}
Eq. \ref{laplacediag} becomes
\begin{eqnarray}
\int_0^{+\infty} d {\cal H}_{ii} {\cal P}_{diag}({\cal H}_{ii}) e^{-t {\cal H}_{ii} } && 
 = e^{- \sqrt{\pi t }   } 
\label{laplacediagcau}
\end{eqnarray}
and the corresponding L\'evy stable law for the diagonal elements ${\cal H}_{ii} $ takes the explicit simple form
\begin{eqnarray}
 {\cal P}_{diag}({\cal H}_{ii}) = \frac{ e^{- \frac{\pi}{4 {\cal H}_{ii} }} }{ 2 {\cal H}_{ii}^{\frac{3}{2}}}
\label{levydiagcau}
\end{eqnarray}

The probability distribution of the product $y_k \equiv H_{1k} H_{2k} $ (Eq \ref{yk}) reads (instead of Eq. \ref{levyyk})
\begin{eqnarray}
R(y) = \frac{ a_N^2 \ln \left( \frac{y^2}{a_N^4} \right) }{\pi^2(y^2-a_N^2) } \oppropto_{ \vert y \vert \to +\infty} 
 \frac{ \ln \left(  \vert y \vert N^2  \right) }{ 2 N^2  y^2}
\label{cauyyk}
\end{eqnarray}
and the probability distribution of the off-diagonal elements ${\cal H}_{ij} $ displays the tail (Eq. \ref{tailoff})
\begin{eqnarray}
 {\cal P}_{off}({\cal H}_{ij}) \opsimeq_{{\cal H}_{ij} \to +\infty}
 \mu^2  
  \frac{ \ln \left(  \vert {\cal H}_{ij} \vert N  \right)}{ 2 N  \vert {\cal H}_{ij} \vert^{1+\mu} }
\label{tailoffcau}
\end{eqnarray}

\subsection{ Properties of the hybridization ratios  }

For $\mu=1$ and $\alpha_0=0$, the typical hybridization ratio 
decays as (Eq \ref{ri0jtyp})
\begin{eqnarray}
R_{i_0j}^{typ}   \propto  N^{-1 }
\label{ri0jtypcau}
\end{eqnarray}
while the hybridization ratio with the site $j_{max}$ having the biggest off-diagonal coupling (Eq. \ref{ri0offmax})
\begin{eqnarray}
R_{i_0j_{max}}   \propto  O(1)
\label{ri0offmaxcau}
\end{eqnarray}
and 
the hybridization ratio with the consecutive energy level then (Eq \ref{rationext})
\begin{eqnarray}
R_{i_0j}^{next}   \propto  O(1)
\label{rationextcau}
\end{eqnarray}
both remain finite.

\subsection{ Probability distribution of $u_q(j) \equiv \vert  {\cal H}_{i_0j} \vert ^{2q}$  }

Eq \ref{distriuq} reads 
\begin{eqnarray}
U_q(u_q) \oppropto_{u_q \to +\infty}  
  \frac{ \ln u_q  }{   N  u_q^{1+\frac{1}{2q} } } 
\label{distriuqcau}
\end{eqnarray}
with the typical value (Eq. \ref{uqtyp})
\begin{eqnarray}
 u_q^{typ} \propto N^{-2q}
\label{uqtypcau}
\end{eqnarray}
and the maximum value (Eq. \ref{uqmax})
\begin{eqnarray}
\opmax_{j \ne i_0} ( u_q(j) ) \propto O(1)
\label{uqmaxcau}
\end{eqnarray}

\subsection{Probability distribution of
 $v_q(j) \equiv \vert {\cal H}_{jj}-{\cal H}_{i_0i_0} \vert ^{-2q}$  }

Eq \ref{distrivq} reads
\begin{eqnarray}
V_q(v_q) \opsimeq_{v_q \to +\infty} \frac{ 1  }{   v_q^{1+\frac{1}{2q}}}
\label{distrivqcau}
\end{eqnarray}
with the typical value (Eq. \ref{vqtyp})
\begin{eqnarray}
 v_q^{typ} \propto O(1)
\label{vqtypcau}
\end{eqnarray}
and the maximum value (Eq. \ref{vqmax})
\begin{eqnarray}
\opmax_{j \ne i_0} ( v_q(j) ) \propto N^{2q}
\label{vqmaxcau}
\end{eqnarray}

\subsection{Probability distribution of
 $w_q(j) \equiv u_q(j) v_q(j)$  }

Here Eqs \ref{distriuqcau} and \ref{distrivqcau} have exactly the same tail exponent,
and one obtains that the probability distribution $W_q$ of $w_q$ displays the tail
\begin{eqnarray}
W_q(w_q) \oppropto_{w_q \to +\infty}  
  \frac{ (\ln w_q)^2  }{   N  w_q^{1+\frac{1}{2q} } } 
\label{distriwqcau}
\end{eqnarray}
so that the maximal value is finite
\begin{eqnarray}
\opmax_{j \ne i_0} ( w_q(j) ) \propto O(1)
\label{wqmaxcau}
\end{eqnarray}
while the typical value decays as
\begin{eqnarray}
w_q^{typ}  \propto  N^{-2q }
\label{wqtypcau}
\end{eqnarray}

\subsection{Probability distribution of  $\Sigma_q$  }

(i) for big indices $q>\frac{1}{2} $, where
the exponent $\frac{1}{2q}  $ in Eq. \ref{distriwqcau}
is smaller than unity, $\Sigma_q$ is a finite random variable
displaying the tail
\begin{eqnarray}
\Pi_q^{sing}(\Sigma_q) \opsimeq_{\Sigma_q \to +\infty}  
 \frac{(\ln \Sigma_q)^2  }{\Sigma_q^{1+\frac{1}{2q}}}
\label{distrisigmaqbigmu}
\end{eqnarray}

(ii) for small indices $0<q<\frac{1}{2} $, 
the exponent $\frac{1}{2q}  $ in Eq. \ref{distriwqcau}
is bigger than unity so that the sum $\Sigma_q$ scales as
\begin{eqnarray}
\Sigma_{q}  \simeq N  w_q^{typ} && \propto   N^{1-2q } 
\label{sqtypqsmallcau}
\end{eqnarray}

\subsection{Multifractality of the IPR $Y_q$  }

 The generalized dimensions of Eq. \ref{multif} then reads
\begin{eqnarray}
D^{criti)}(q) && =  0   \  \ \  \ \ \  \ \ \ \ \ \  \ \ \ \ \  \ \ \ {\rm for } \ \ \ q> \frac{1}{2} 
\nonumber \\
D^{criti}(q) && =  \frac{1- 2q}{1-q}  \ \ \  \ \ \ {\rm for } \ \ \ 0<q< \frac{1}{2} 
\label{multifmubig}
\end{eqnarray}
which corresponds to the following linear multifractal spectrum : 
the number of sites with the density $\psi^2(i) \propto N^{-\gamma}  $ scales as 
\begin{eqnarray}
{\cal N}^{criti} ( \psi^2(i) \propto N^{-\gamma}   )
&& \propto   N^{\frac{\gamma}{2}}  \theta (0 \leq \gamma \leq 2)
\label{strongmultif}
\end{eqnarray}

This spectrum is well-known as the 'Strong Multifractality' critical spectrum \cite{mirlin_fyodorov,mirlin_four}
which is the farthest possible from the ergodic phase that corresponds to a delta function at $\gamma=1$ 
and the closest possible from the localized phase with its Poisson statistics for the energy levels.
It appears in particular at Anderson Localization Transition
in the limit of infinite dimension $d \to +\infty$ \cite{mirlinrevue}
or in long-ranged power-law hoppings in one-dimension 
 \cite{levitov1,levitov2,levitov3,levitov4,mirlin_evers,fyodorov,
fyodorovrigorous,oleg1,oleg2,oleg3,oleg4,oleg5,oleg6,olivier_per,olivier_strong,olivier_conjecture,us_strongmultif},
or in the matrix model of \cite{kravtsov_nonergomf}
Recently, it has been also found for the Many-Body-Localization transition \cite{c_mblrgeigen,c_mblperturb}.
Although these various models seem very different from a physical point of view, they all have the technical property that the localized eigenvectors are
not exponentially localized but only power-law localized with respect to the size $N$ of the Hilbert space.
Then, as explained in the introduction, there is already some multifractality within the localized phase (Eq \ref{locmultif}),
so that the multifractality at the critical point is reached as the limit of the localized multifractality at the limit of stability of the localized phase :
for instance in the present case, Eq. \ref{strongmultif} corresponds to the limit of the spectrum of Eq. \ref{levymultifR} as $\mu \to 1$.
This continuity between the localized multifractal spectrum and the critical 'Strong Multifractality' 
has already been found in the matrix model of \cite{kravtsov_nonergomf}
and for Many-Body-Localization models \cite{c_mblrgeigen,c_mblperturb}.

\section{Conclusion }

\label{sec_conclusion}

To analyze the localization properties of a random L\'evy matrix $H$, 
we have proposed to consider the square ${\cal H} = H^2$ in order to produce a strong hierarchy between the diagonal elements ${\cal H}_{ii}$ 
and the off-diagonal elements ${\cal H}_{i<i}$. The off-diagonal elements can be then taken into account  via the standard first order perturbation theory 
of quantum mechanics as long as the hybridization ratios do not diverge in the thermodynamic limit.
This strong disorder perturbative expansion remains consistent within the localized phase 
and allows to study the Inverse Participation Ratios (I.P.R.)  $Y_q$ of the eigenvectors as a function of the continuous parameter $0<q<+\infty$.

 In the region $0<\mu<1$, we have found that all eigenvectors are localized but display some multifractality : the IPR are finite above some threshold $q>q_c$ but diverge in the region $0<q<q_c$ near the origin. 

In the region $1<\mu<2$, we have obtained that only the sub-extensive fraction $N^{\frac{3}{2+\mu}}$ of the biggest eigenvalues corresponding to the region $\vert E \vert \geq N^{\frac{(\mu-1)}{\mu(2+\mu)}} $ remains localized, while the extensive number of other states of smaller energy are delocalized. 

For the extensive number of finite eigenvalues, the localization/delocalization transition thus takes place 
at the critical value corresponding to Cauchy matrices : the Inverse Participation Ratios $Y_q$ of the corresponding critical eigenstates follow the Strong-Multifractality Spectrum  which is well-known in various other Localization problems in spaces of effective infinite dimensionality.

\end{document}